\documentclass[floats,aps,amssymb,nofootinbib]{revtex4}
\usepackage{amssymb}
\usepackage{graphics}
\usepackage{amsmath}
\usepackage{amsfonts}
\usepackage{bm}
\usepackage[]{latexsym}
\usepackage{float}
\usepackage{graphicx}

\newcommand{\be}{\begin{equation}}\newcommand{\ee}{\end{equation}}
\newcommand{\bea}{\begin{eqnarray}}\newcommand{\eea}{\end{eqnarray}}
\newcommand{\brr}{\begin{array}}\newcommand{\err}{\end{array}}
\newcommand{\bit}{\begin{itemize}}\newcommand{\eit}{\end{itemize}}
\newcommand{\ben}{\begin{enumerate}}\newcommand{\een}{\end{enumerate}}

\newcommand{\ba}{\begin{array}}
\newcommand{\ea}{\end{array}}

\def\lf{\left}

\def\pa{\partial}

\def\ri{\right}
\def\al{\alpha}

\def\si{\sigma}

\def\1{{_{1}}}\def\2{{_{2}}}

\def\noHe0{:\;\!\!\;\!\!:H_e(0):\;\!\!\;\!\!:}
\def\noHm0{:\;\!\!\;\!\!:H_\mu(0):\;\!\!\;\!\!:}

\def\lf{\left}

\def\pa{\partial}

\def\ri{\right}

\def\al{\alpha}

\def\si{\sigma}

\def\1{{_{1}}}\def\2{{_{2}}}

\begin{document}
\title{Equivalence principle violation from large scale structure}

\author{Luciano Petruzziello\footnote{lupetruzziello@unisa.it}$^{\hspace{0.3mm}1,2}$} \affiliation
{$^1$Dipartimento di Ingegneria Industriale, Universit\`a degli Studi di Salerno, Via Giovanni Paolo II, 132 I-84084 Fisciano (SA), Italy. \\
$^2$INFN, Sezione di Napoli, Gruppo collegato di Salerno, Italy.}

\date{\today}
\def\be{\begin{equation}}
\def\ee{\end{equation}}
\def\al{\alpha}
\def\bea{\begin{eqnarray}}
\def\eea{\end{eqnarray}}

\begin{abstract}
We explore the interplay between the equivalence principle and a generalization of the Heisenberg uncertainty relations known as extended uncertainty principle, that comprises the effects of spacetime curvature at large distances. Specifically, we observe that, when the modified uncertainty relations hold, the weak formulation of the equivalence principle is violated, since the inertial mass of quantum systems becomes position-dependent whilst the gravitational mass is left untouched. To obtain the above result, spinor and scalar fields are separately analyzed by considering the non-relativistic limit of the Dirac and the Klein-Gordon equations in the presence of the extended uncertainty principle. In both scenarios, it is found that the ratio between the inertial and the gravitational mass is the same.
\end{abstract}

\vskip -1.0 truecm 

\maketitle

\section{Introduction}

One of the cornerstones of the celebrated general theory of relativity is undoubtedly represented by the equivalence principle (EP). According to Einstein, the (locally) indistinguishable equivalence between a gravitational field and an acceleration of the reference frame has been the main guiding light for the development of a relativistic treatment of gravitation. Ever since this archetypical notion was introduced, the equivalence principle has been thoroughly investigated from several standpoints, thus allowing for a plethora of distinct, brand-new interpretations. Indeed, when the equivalence principle is addressed nowadays there is a certain degree of ambiguity, as it may come in different formulations, each of which has its own requirements and range of validity. For a complete and detailed overview on this topic, the interested reader can consult Refs.~\cite{libe,new}. 

For the purpose of the present paper, we restrict our attention to the weak equivalence principle, which establishes the equality between the inertial and the gravitational mass, henceforth labeled as $m_i$ and $m_g$, respectively. However, the aforementioned equality can be heavily influenced by the physical system and environment under consideration, as under specific circumstances the two masses may no longer have the same magnitude. For instance, when studying an electrically charged particle that interacts with a thermal bath of photons, radiative corrections affect $m_i$ and $m_g$ separately and unevenly, thereby giving rise to a violation of EP in its weak formulation~\cite{new,don,our}. The same occurrence can be found in the analysis of non-relativistic neutrinos, where the flavor mixing is responsible for a redefinition of the inertial mass, which thus becomes distinguishable from $m_g$~\cite{neutrino}. On a broader scale, the interplay between neutrino physics and EP violation is much deeper than the above feature, as the vast literature on the subject confirms~\cite{wlw,ms,wwg,sck,nmf,sah,ash,bha,ours,gn,hl}.

Here, we investigate a different scenario where EP violation is manifest. Specifically, we consider a relativistic quantum mechanical framework whose underlying Heisenberg uncertainty principle (HUP) is modified so as to accommodate the existence of a minimal momentum scale~\cite{kempf,kempf2,cbl,mz,bambi,oy,mignemi,symm}; a similar generalization goes by the name of extended uncertainty principle (EUP). The justification for the above deformation sinks its roots in cosmological arguments, since the curvature of spacetime at large distances is so extreme that there is a fundamental limit to the resolution with which momentum can be described~\cite{kempf,kempf2,mz}. As an additional piece of evidence for this observation, it is no coincidence that, in accordance with several findings~\cite{bambi,oy,mignemi}, a well-known formulation of the EUP is based upon the employment of the cosmological constant. Nevertheless, it has also been shown~\cite{schur,schur2,fab,fab2,fab3} that another version of EUP is deduced from a straightforward application of quantum mechanical laws in a general relativistic setting, thus letting the deviation from HUP depend exclusively on the intrinsic aspects of the spacetime under examination. In this work, we will analyze the most common setting, according to which the characteristic parameter associated to EUP must be fixed by requiring consistency with experimental data. In this regard, it is worth mentioning that the most stringent bounds on the EUP parameter are discussed in Refs.~\cite{mz,symm}.

In order to comply with the aforementioned purposes, the paper is structured as follows: in Sec.~II, the main features of the extended uncertainty principle are summarized, focusing in particular on the correction it entails at the level of the momentum operator. Section~III is devoted to the non-relativistic limit of the Dirac equation with the modified momentum, whilst the same study in conjunction with the Klein-Gordon equation is the topic of Sec.~IV. Finally, Sec.~V contains concluding remarks and future perspectives. 

Throughout the whole manuscript, we use the natural units $\hbar=c=1$ but keep the speed of light explicit only in Sec.~IV, since the non-relativistic regime for a scalar field can be readily derived from a perturbative treatment of the Klein-Gordon equation in powers of $c$ (as seen in Ref.~\cite{kiefer}).


\section{Extended uncertainty principle}

In simple terms, the Heisenberg uncertainty principle states that arbitrarily small spatial distances can be resolved with sufficiently high-energy probes. To mathematically express this concept, we know that
\be\label{hup}
\Delta x\Delta p\geq\frac{1}{2}\,.
\ee
However, if a minimal momentum scale is accounted for, this is no longer the case; indeed, the previous equation must be suitably modified. Drawing inspiration from Refs.~\cite{kempf,kempf2,cbl,mz,bambi,oy,mignemi}, we argue that the influence of spacetime curvature at large distances induces a position-dependent correction into Eq.~\eqref{hup}, which then becomes
\be\label{eup}
\Delta X\Delta P\geq\frac{1}{2}\lf(1+\al\Delta X^2\ri)\,,
\ee
where $\al$ is the so-called deformation parameter and we assume to work in the regime in which the dimensionless quantity $\al\Delta X^2$ can be regarded as a perturbation. Since the deformation parameter is the inverse of a squared length, the close relation between $\al$ and the cosmological constant $\Lambda$ can be easily inferred. As a matter of fact, it was shown in Ref.~\cite{park} that $\al=\pm\Lambda/3$, where the minus (plus) sign is associated to a (anti-)de Sitter spacetime. Therefore, in compliance with the experimental observations, we deduce that $\al<0$. Nonetheless, as already anticipated, we analyze the more general scenario where $\al$ may not be necessarily related to $\Lambda$, in line with Refs.~\cite{kempf,kempf2,mz,symm}.
 
At this point, from Eq.~\eqref{eup} it is straightforward to compute the value of the minimal uncertainty in momentum, which turns out to be $\Delta P_{\mathrm{min}}=\sqrt{|\al|}$. Furthermore, from the same equation, it is also possible to trace back to a general expression for the deformed commutator between the position and momentum operators. Indeed, starting from
\be\label{comm}
\lf[X,P\ri]=i\lf(1+\al X^2\ri)\,,
\ee
one can make use of the Robertson-Schr\"odinger prescription to deduce Eq.~\eqref{eup}. 

The deformed canonical commutator plays a relevant role, as it provides an easy representation of $X$ and $P$ in terms of the auxiliary operators $x$ and $p$ for which the usual relation $\lf[x,p\ri]=i$ holds. To see this, it is sufficient to replace the following identities
\be\label{op}
X=x\,, \qquad P=\lf(1+\al x^2\ri)p-i\al x\,,
\ee
into Eq.~\eqref{comm} and verify that it is always satisfied. The second term in the definition of the physical momentum has been added to make the operator self-adjoint.

The identification~\eqref{op} is valid in one spatial dimension and as long as all the orders higher than and up to $\mathcal{O}(\al^2)$ are neglected. Should we perform the full expansion with more than one spatial dimension, we would have to deal with an emerging non-commutativity for different momentum components, because $\lf[P_i,P_j\ri]\propto\mathcal{O}(\al^2)$. However, for the purposes of the present paper, we can safely restrict the attention to the lowest-order terms in the deformation parameter without harming the generality of our results. For further details on higher-order EUP, see Refs.~\cite{kempf,kempf2,mignemi}.

Before concluding this Section, we point out that calculations will be carried out in the position representation; hence, by generalizing Eq.~\eqref{op} to higher dimensions, we obtain
\be\label{op2}
X_i=x_i\,, \qquad P_i=-i(1+\al x^2)\pa_i-i\al x_i\,, 
\ee
with $x^2=\sum_ix_i^2$.


\section{Equivalence principle violation for spinor fields}

In order to approach the modified Dirac equation, one has to modify
\be\label{deq}
\lf(i\gamma^\mu\pa_\mu-m\ri)\psi=0\,,
\ee
by properly generalizing the auxiliary momentum operator to the physical one in the context of EUP. This is achieved by exploiting the relation~\eqref{op2} to cast the extended version of the Dirac equation as
\be\label{mdeq}
i\gamma^0\pa_t\psi=-i\lf(1+\al x^2\ri)\vec{\gamma}\cdot\vec{\nabla}\psi-i\al\vec{\gamma}\cdot\vec{x}\,\psi+m\psi\,.
\ee
A similar field equation has already been addressed in Ref.~\cite{gine}, where the authors discuss a completely unrelated issue. 

The gamma matrices appearing in~\eqref{mdeq} can be written in the Dirac representation, namely 
\be\label{gamma}
\gamma^0=\begin{pmatrix} \textbf{1} & 0 \\ 0 & -\textbf{1} \end{pmatrix}\,, \qquad \vec{\gamma}=\begin{pmatrix} 0 & \vec{\si} \\ -\vec{\si} & 0 \end{pmatrix}\,,
\ee
where $\textbf{1}$ denotes the $2\times2$ identity matrix and $\vec{\si}=\lf(\si_x,\si_y,\si_z\ri)$ the Pauli matrices.

To provide a comprehensive analysis of the subject, one shall incorporate the effects of an external gravitational field in such a way to have access to both the inertial and the gravitational mass from the non-relativistic limit. Indeed, along the line of Refs.~\cite{don}, it is known that the two different masses can only be distinguished within the non-relativistic Schr\"odinger equation for the particle sector. This identification is achieved by looking at the kinetic operator to deduce $m_i$ and the interaction with the gravitational potential to derive $m_g$. 

In order to study the dynamics of a spinor field in the presence of gravity, Eq.~\eqref{deq} must be modified by introducing the spin connection $\Gamma_\mu$ and vierbein fields $e^\mu_a$ as follows:
\be\label{gdeq}
\Big[i\gamma^ae^\mu_a\lf(\pa_\mu+\Gamma_\mu\ri)-m\Big]\psi=0\,.
\ee
The nature of $\Gamma_\mu$ and $e^\mu_a$ is closely intertwined with the shape of the metric tensor; for a detailed review, see Ref.~\cite{gravitation}. Here, since we seek the non-relativistic limit and an easy-to-interpret interaction with the gravitational field, we require that the metric tensor describes a weak-field approximation of the Schwarzschild solution in isotropic coordinates, i.e.,
\be\label{metric}
ds^2=\lf(1+2\phi\ri)dt^2-\lf(1-2\phi\ri)d\textbf{r}^2\,,
\ee
where the potential $\phi=-GM/r$ is assumed to be a small quantity. From Eq.~\eqref{metric}, one can promptly compute the spin connection and the vierbeins; a simple calculation (which can be found in Ref.~\cite{neutrino}) shows that
\be\label{spinc}
\Gamma_\mu=\frac{1}{8}\lf[\gamma^a,\gamma^b\ri]e^\lambda_a\lf(\eta_{\mu\lambda}\pa_\rho\phi-\eta_{\mu\rho}\pa_\lambda\phi\ri)e^\rho_b\,,
\ee
\be\label{tetrad}
e^t_0=1-\phi\,, \qquad e^x_1=e^y_2=e^z_3=1+\phi\,.
\ee
At this point, one can further require that the spatial gradients of the gravitational potential are negligible, so that $\pa_i\phi\approx0$, $\forall i$. Although such a choice does not affect the validity of our results, it turns out to be useful, as it significantly streamlines the intermediate steps. As a matter of fact, by virtue of this assumption, the spin connection~\eqref{spinc} is identically vanishing, which means that the influence of gravity only enters the Dirac equation through the vierbein fields. In light of this, the extension of Eq.~\eqref{mdeq} which is inclusive of the gravitational interaction reads
\be\label{mdeq2}
i\gamma^0\lf(1-\phi\ri)\pa_t\psi=-i\lf(1+\al x^2+\phi\ri)\vec{\gamma}\cdot\vec{\nabla}\psi-i\al\vec{\gamma}\cdot\vec{x}\,\psi+m\psi\,.
\ee
We can now proceed by separating the Dirac spinor into the upper and lower components
\be\label{spinor}
\psi=\begin{pmatrix} \varphi \\ \chi \end{pmatrix}\,,
\ee
so as to explicitly deal with the equations for $\varphi$ and $\chi$ independently. In so doing, one has
\be\label{varphi}
i\lf(1-\phi\ri)\pa_t\varphi=-i\lf(1+\al x^2+\phi\ri)\vec{\si}\cdot\vec{\nabla}\chi-i\al\vec{\si}\cdot\vec{x}\,\chi+m\varphi\,,
\ee
\be\label{chi}
-i\lf(1-\phi\ri)\pa_t\chi=i\lf(1+\al x^2+\phi\ri)\vec{\si}\cdot\vec{\nabla}\varphi+i\al\vec{\si}\cdot\vec{x}\,\varphi+m\chi\,.
\ee
In order to reach the non-relativistic regime, we first must note that, in this limit, the most important contribution to the energy comes from the rest mass. Therefore, if we single out a fast oscillating factor $\exp(-imt)$ from the Dirac spinor, we have that~\cite{neutrino,drell} 
\be\label{nrl}
\psi=e^{-imt}\tilde{\psi}\,,
\ee
but since $\tilde{\psi}$ still depends on the time coordinate, it is immediate to realize $|i\pa_t\tilde{\psi}|\ll|m\tilde{\psi}|$, which provides a non-relativistic approximation~\cite{drell}. In light of this result, it is possible to cast $\varphi$ and $\chi$ in terms of $\tilde{\varphi}$ and $\tilde{\chi}$ and enforce the above condition; the outcome of such a step yields
\be\label{varphi2}
i\pa_t\tilde{\varphi}=m\phi\,\tilde{\varphi}-i\lf(1+\al x^2+\phi\ri)\vec{\si}\cdot\vec{\nabla}\tilde{\chi}-i\al\vec{\si}\cdot\vec{x}\,\tilde{\chi}\,,
\ee
\be\label{chi2}
\lf(\phi-2\ri)m\tilde{\chi}=i\lf(1+\al x^2+\phi\ri)\vec{\si}\cdot\vec{\nabla}\tilde{\varphi}+i\al\vec{\si}\cdot\vec{x}\,\tilde{\varphi}\,.
\ee
The latter equation can be employed to find an explicit dependence of $\tilde{\chi}$ on $\tilde{\varphi}$. By recalling the weak-field approximation, we obtain 
\be\label{chi3}
\tilde{\chi}=-\frac{i}{2m}\lf(1+\al x^2+\frac{3}{2}\phi\ri)\vec{\si}\cdot\vec{\nabla}\tilde{\varphi}-\frac{i\al}{2m}\vec{\si}\cdot\vec{x}\,\tilde{\varphi}\,.
\ee
As it also occurs for the standard Dirac equation, in the non-relativistic regime the lower component $\chi$ is sub-dominant with respect to the upper component $\varphi$~\cite{drell}. Hence, we can focus on Eq.~\eqref{varphi2} only and derive the Schr\"odinger-like equation for the particle sector. To this aim, the expression~\eqref{chi3} must be substituted in \eqref{varphi2}, so as to get
\be\label{find1}
i\pa_t\tilde{\varphi}=-\lf(1+2\al x^2\ri)\frac{\nabla^2\tilde{\varphi}}{2m}+m\phi\,\tilde{\varphi}-\frac{5}{4m}\phi\,\nabla^2\tilde{\varphi}-\frac{2\al}{m}\lf(\vec{\si}\cdot\vec{x}\ri)\vec{\si}\cdot\vec{\nabla}\tilde{\varphi}-\frac{3\al}{2m}\tilde{\varphi}\,.
\ee
A further manipulation of Eq.~\eqref{find1} allows us to reach the final expression, where the non-relativistic Schr\"odinger equation associated with $\tilde{\varphi}$ is manifest
\be\label{find}
i\pa_t\tilde{\varphi}=-\lf(1+2\al x^2\ri)\frac{\nabla^2\tilde{\varphi}}{2m}+m\phi\,\tilde{\varphi}-\frac{5}{4m}\phi\,\nabla^2\tilde{\varphi}-\frac{2i\al}{m}\vec{\si}\cdot\lf(\vec{x}\wedge\vec{\nabla}\tilde{\varphi}\ri)-\frac{2\al}{m}\vec{x}\cdot\vec{\nabla}\tilde{\varphi}-\frac{3\al}{2m}\tilde{\varphi}\,.
\ee
The relevant factors to be considered are the first two appearing in the r.h.s., which represent the kinetic operator and the interaction with the external gravitational potential, respectively. The third term can be interpreted as a post-Newtonian correction, whilst the fourth one may be viewed as an EUP-induced spin-orbit interaction. The last two quantities are distinctive of the EUP quantum mechanical treatment, but the study of their implications lies beyond the purpose of the present paper.

Now, we need to extrapolate the shape of the inertial and gravitational mass connected with the spinor field; concerning $m_i$, its definition can be directly deduced from the kinetic operator, and turns out to be 
\be\label{mi}
m_i=m\lf(1-2\al x^2\ri)\,.
\ee 
On the other hand, the gravitational mass is easily recognizable from the gravitational interaction, which tells us that $m_g=m$. Therefore, $m_i$ and $m_g$ do not undergo the same redefinition, which in turn signals a violation of the weak equivalence principle. This violation can be quantified by the ratio
\be\label{ratio}
\frac{m_g}{m_i}=1+2\al x^2\,,
\ee
which smoothly recovers the fulfillment of EP as long as $\al\to0$.

In the next Section, we exhibit that a similar feature emerges not only in the context of spinor fields, but also for scalar particles. Remarkably, the degree of EP violation will be quantified by the same function as in Eq.~\eqref{ratio}, thus suggesting that the present achievement does not depend on the nature of the system under investigation.


\section{Equivalence principle violation for scalar fields}

The starting point for the analysis of scalar fields is the Klein-Gordon equation
\be\label{kg}
\lf(\pa_\mu\pa^\mu+m^2c^2\ri)\psi=0\,,
\ee
where the speed of light is kept explicit for later convenience. In the presence of gravity, the above equation is generalized as~\cite{gravitation}
\be\label{gkg}
\frac{1}{\sqrt{-g}}\pa_\mu\lf(\sqrt{-g}g^{\mu\nu}\pa_\nu\psi\ri)+m^2c^2\psi=0\,,
\ee
where $g=\mathrm{det}(g_{\mu\nu})$. In order to establish a faithful comparison with the spinor case, $g_{\mu\nu}$ must be the one described by Eq.~\eqref{metric}. However, by restoring $c$ appropriately, the line element~\eqref{metric} now reads
\be\label{metric2}
ds^2=\lf(1+2\frac{\phi}{c^2}\ri)c^2dt^2-\lf(1-2\frac{\phi}{c^2}\ri)d\textbf{r}^2\,,
\ee
with $\phi$ left untouched.

By embedding the above physical setting in the framework of EUP and neglecting contributions that go like $\mathcal{O}(\al^2)$, Eq.~\eqref{gkg} becomes
\be\label{gkg2}
\frac{1}{c^2}\lf(1-2\frac{\phi}{c^2}\ri)\pa^2_t\psi-\lf(1+2\al x^2+2\frac{\phi}{c^2}\ri)\nabla^2\psi-4\al\vec{x}\cdot\vec{\nabla}\psi-3\al\psi+m^2c^2\psi=0\,.
\ee
Equation~\eqref{gkg2} has been thoroughly investigated in Ref.~\cite{chung}, where the scalar particle couples to an external magnetic field rather than the gravitational one.

In order to consider the non-relativistic limit, we can perform an expansion in powers of $c$. In accordance with Ref.~\cite{kiefer}, a possible solution for the Klein-Gordon equation can be written as
\be\label{sol}
\psi\lf(\vec{x},t\ri)=e^{i\varphi\lf(\vec{x},t\ri)}\,,
\ee
where $\varphi$ is assumed to depend on $c$ as
\be\label{exp}
\varphi=c^2\varphi_0+\varphi_1+c^{-2}\varphi_2+\dots\,.
\ee
The series is infinite, but for our purposes we can focus the attention on the three contributions appearing in Eq.~\eqref{exp}. Clearly, the above working hypothesis entails that any derivative acting on $\psi$ results in the emergence of further powers of $c$ in Eq.~\eqref{gkg2}. Specifically, by denoting with $\pa$ a generic derivative (which can be either spatial or temporal), one has
\be\label{fd}
\pa\psi=i\lf(c^2\pa\varphi_0+\pa\varphi_1+c^{-2}\pa\varphi_2\ri)e^{i\varphi}\,,
\ee
whilst second derivatives imply
\be\label{sd}
\pa^2\psi=\lf\{i\Big[c^2\pa^2\varphi_0+\pa^2\varphi_1+c^{-2}\pa^2\varphi_2\Big]-\lf[c^4\lf(\pa\varphi_0\ri)^2+2c^2\pa\varphi_0\pa\varphi_1+2\pa\varphi_0\pa\varphi_2+\lf(\pa\varphi_1\ri)^2+\frac{2}{c^2}\pa\varphi_1\pa\varphi_2+\frac{\lf(\pa\varphi_2\ri)^2}{c^4}\ri]\ri\}e^{i\varphi}\,.
\ee
We can now apply the aforementioned scheme to Eq.~\eqref{gkg2} in order to write a Klein-Gordon equation which explicitly depends on $\varphi_0$, $\varphi_1$ and $\varphi_2$. To achieve the sought non-relativistic limit, the analysis must be truncated at $c^0$ starting from the highest order in the expansion parameter, which turns out to be $c^4$. Despite the seeming complexity, computations are instead straightforward, because in order to satisfy Eq.~\eqref{gkg2} it must necessarily be imposed that all the functions in front of any power of $c$ must vanish identically, order by order~\cite{kiefer}.  

The first step is to look at the order $c^4$, whose contribution yields
\be\label{c4}
\lf(1+2\al x^2\ri)(\vec{\nabla}\varphi_0)^2=0\,,
\ee
which tells that $\varphi_0$ is a function of time only. The explicit expression for $\varphi_0$ can be deduced from the order $c^2$, which reads 
\be\label{c2}
-\lf(\pa_t\varphi_0\ri)^2+m^2=0\,.
\ee
Hence, up to $c^2$ the solutions to the Klein-Gordon equation are simply $\psi=\exp(\pm i m c^2t)$, which represent the particle/antiparticle wave functions with $E$ being the rest energy $m c^2$. In the non-relativistic regime, no particle creation occurs, so the two scenarios can be treated separately; however, since they would give the exactly same outcome at the next order, we can arbitrarily choose to work with only one of them. Thus, by selecting the minus sign (i.e., particle wave function) and moving on to the order $c^0$, we obtain
\be\label{c0}
2m\pa_t\varphi_1-i\lf(1+2\al x^2\ri)\nabla^2\varphi_1-4i\al\vec{x}\cdot\vec{\nabla}\varphi_1-3\al+2m^2\phi+\lf(1+2\al x^2\ri)\lf(\vec{\nabla}\varphi_1\ri)^2=0\,.
\ee
By introducing $\psi_1=\exp(i\varphi_1)$, the previous equation can be cast in a Schr\"odinger-like form, that is
\be\label{finkg}
i\pa_t\psi_1=-\lf(1+2\al x^2\ri)\frac{\nabla^2\psi_1}{2m}+m\phi\psi_1-\frac{2\al}{m}\vec{x}\cdot\vec{\nabla}\psi_1-\frac{3\al}{2m}\psi_1\,.
\ee
Higher-order terms in $c$ would only have added relativistic corrections to Eq.~\eqref{finkg}, which thereby represents the non-relativistic limit of Eq.~\eqref{gkg2}. 

As for Eq.~\eqref{find}, we recover the two characteristic terms which are reminiscent of the minimal momentum scale (namely, the last two quantities of the r.h.s.). Furthermore, from the first two factors of Eq.~\eqref{finkg} the definitions of $m_i$ and $m_g$ can be immediately deduced; it turns out that the two masses are identical to the ones already found for the spinor field. In light of this, we conclude that the equivalence principle does not hold for scalar fields in the presence of the extended uncertainty principle, and the degree of its violation is precisely equal to the one evaluated with the Dirac equation, i.e., Eq.~\eqref{ratio}.


\section{Concluding remarks}

In this paper, we have shown how the existence of a minimal momentum scale induces a violation of the weak equivalence principle. To demonstrate the generality of our statement, we have separately considered spinor and scalar fields to verify that, in both cases, the degree of EP violation is still the same, thus suggesting that a similar occurrence may not depend on the nature of the examined quantum system. To unambiguously recognize the magnitude of the inertial and gravitational mass, we have performed a non-relativistic limit onto the Dirac and the Klein-Gordon equations, which have been written so as to simultaneously account for the presence of EUP and a weak and static gravitational field. In so doing, we have then identified $m_i$ with the mass term appearing in the kinetic part of the Schr\"odinger Hamiltonian and $m_g$ with the factor that multiplies the gravitational potential $\phi$. The same strategy has already been used in different frameworks with success~\cite{don,neutrino}, and it still represents the most reliable option, given that in the fully relativistic regime such an identification is a challenging task. 

On a final note, we want to remark that, because of the physical origins of the extended uncertainty principle, the implications of the achieved result become relevant at cosmological scales. This concept can be quantified by comparing the EP-violating factor $2\al x^2$ with the available experimental constraint for the term $|m_g-m_i|/m_i$. As a matter of fact, by recalling that the above ratio is bounded by a factor $10^{-14}$~\cite{neutrino}, we immediately obtain 
\be\label{phys}
x\simeq\frac{10^{-7}}{\sqrt{|\al|}}\,.
\ee
Since the magnitude of $\al$ is expected to be extremely small, the characteristic length scale $x$ at which the EP violation becomes significant is huge. By resorting to the formulation of EUP where $|\al|\simeq\Lambda\simeq10^{-52}$ m$^{-2}$, a straightforward calculation would yield $x_\Lambda\simeq10^{19}$ m, which is roughly $10^3$ light years. On the other hand, a more conservative approach that relies on experimental bounds for the EUP deformation parameter provides $\sqrt{|\al|}\lesssim10^{-12}\,\,\mathrm{m}^{-1}$~\cite{symm}, which instead would give a lower bound, that is $x_{\mathrm{min}}\gtrsim10^5\,\,\mathrm{m}$. Thence, the most promising phenomenological perspectives for the EUP-induced equivalence principle violation lies in the interval $[x_{\mathrm{min}},x_\Lambda]$. To unravel further pieces of information, more work is inevitably required along this direction.

\section*{Acknowledgement}

The author acknowledges networking support by the COST Action CA18108.


\begin{thebibliography}{0}
\section*{References}
\smallskip
 
\bibitem{libe}
E.~Di Casola, S.~Liberati and S.~Sonego,
Am. J. Phys. \textbf{83}, 39 (2015).

\bibitem{new}
M.~Blasone, S.~Capozziello, G.~Lambiase and L.~Petruzziello,
arXiv:2112.08480 [gr-qc] (2021), doi: 10.1142/S0219887822500554.

\bibitem{don}
J.~F.~Donoghue, B.~R.~Holstein and R.~W.~Robinett,
Phys. Rev. D \textbf{30}, 2561 (1984);
J.~F.~Donoghue, B.~R.~Holstein and R.~W.~Robinett,
Gen. Rel. Grav. \textbf{17}, 207 (1985).

\bibitem{our}
M.~Blasone, S.~Capozziello, G.~Lambiase and L.~Petruzziello,
Eur. Phys. J. Plus \textbf{134}, 169 (2019).

\bibitem{neutrino}
M.~Blasone, P.~Jizba, G.~Lambiase and L.~Petruzziello,
Phys. Lett. B \textbf{811}, 135883 (2020).

\bibitem{wlw}
Z.~Y. Wang, R.~Y. Liu and X.~Y. Wang, Phys. Rev. Lett. \textbf{116}, 151101 (2016).

\bibitem{ms}
  R.~B.~Mann and U.~Sarkar,
  Phys.\ Rev.\ Lett.\  {\bf 76}, 865 (1996).

\bibitem{wwg}
  J.~J.~Wei, X.~F.~Wu, H.~Gao and P.~Maszaros,
  JCAP {\bf 1608}, 031 (2016).
	
\bibitem{sck}
  I.~I.~Shapiro, C.~C.~Counselman and R.~W.~King,
  Phys.\ Rev.\ Lett.\  {\bf 36}, 555 (1976).	
	
\bibitem{nmf}	
 T.~M.~Niebauer, M.~P.~Mchugh and J.~E.~Faller,
  Phys.\ Rev.\ Lett.\  {\bf 59}, 609 (1987).
	
\bibitem{sah}	
C.~W.~Stubbs, E.~G.~Adelberger, B.~R.~Heckel, W.~F.~Rogers, H.~E.~Swanson, R.~Watanabe, J.~H.~Gundlach and F.~J.~Raab,
  Phys.\ Rev.\ Lett.\  {\bf 62}, 609 (1989).
	
\bibitem{ash}
  E.~G.~Adelberger, C.~W.~Stubbs, B.~R.~Heckel, Y.~Su, H.~E.~Swanson, G.~Smith, J.~H.~Gundlach and W.~F.~Rogers,
  Phys.\ Rev.\ D {\bf 42}, 3267 (1990).
	
\bibitem{bha}	
  S.~Baessler, B.~R.~Heckel, E.~G.~Adelberger, J.~H.~Gundlach, U.~Schmidt and H.~E.~Swanson,
  Phys.\ Rev.\ Lett.\  {\bf 83}, 3585 (1999).
	
\bibitem{ours}
L.~Buoninfante, G.~G.~Luciano, L.~Petruzziello and L.~Smaldone,
  Phys.\ Rev.\ D {\bf 101}, 024016 (2020).	
	
\bibitem{gn}
 A.~M.~Gago, H.~Nunokawa and R.~Zukanovich Funchal,
  Nucl.\ Phys.\ Proc.\ Suppl.\  {\bf 100}, 68 (2001).	
	
\bibitem{hl}
 J.~T.~Pantaleone, A.~Halprin and C.~N.~Leung,
  Phys.\ Rev.\ D {\bf 47}, R4199 (1993);
  A.~Halprin, C.~N.~Leung and J.~T.~Pantaleone,
  Phys.\ Rev.\ D {\bf 53}, 5365 (1996).	
	
\bibitem{kempf}
H.~Hinrichsen and A.~Kempf,
J. Math. Phys. \textbf{37}, 2121 (1996).

\bibitem{kempf2}
A.~Kempf,
J. Math. Phys. \textbf{38}, 1347 (1997).


\bibitem{cbl}
R.~N.~Costa Filho, J.~P.~M.~Braga, J.~H.~S.~Lira and J.~S.~Andrade,
Phys. Lett. B \textbf{755}, 367 (2016).

\bibitem{mz}	
M.~Zarei and B.~Mirza,
Phys. Rev. D \textbf{79}, 125007 (2009).

\bibitem{bambi}
C.~Bambi and F.~R.~Urban,
Class. Quant. Grav. \textbf{25}, 095006 (2008).

\bibitem{oy}
Y.~C.~Ong and Y.~Yao,
Phys. Rev. D \textbf{98}, 126018 (2018).

\bibitem{mignemi}
S.~Mignemi,
Mod. Phys. Lett. A \textbf{25}, 1697 (2010).

\bibitem{symm}
F.~Illuminati, G.~Lambiase and L.~Petruzziello,
Symmetry \textbf{13}, 1854 (2021).

\bibitem{schur}
T.~Schurmann and I.~Hoffmann,
Found. Phys. \textbf{39}, 958 (2009).

\bibitem{schur2}
T.~Sch\"urmann,
Found. Phys. \textbf{48}, 716 (2018).

\bibitem{fab}
M.~P.~Dabrowski and F.~Wagner,
Eur. Phys. J. C \textbf{79}, 716 (2019).

\bibitem{fab2}
M.~P.~Dabrowski and F.~Wagner,
Eur. Phys. J. C \textbf{80}, 676 (2020).

\bibitem{fab3}
L.~Petruzziello and F.~Wagner,
Phys. Rev. D \textbf{103}, 104061 (2021).

\bibitem{kiefer}
C.~Kiefer and T.~P.~Singh,
Phys. Rev. D \textbf{44}, 1067 (1991).

\bibitem{park}
M.~i.~Park,
Phys. Lett. B \textbf{659}, 698 (2008).

\bibitem{gine}
D.~Chemisana, J.~Gin\'e and J.~Madrid,
Phys. Scripta \textbf{96}, 065311 (2021).

\bibitem{gravitation}
C.~W.~Misner, K.~S.~Thorne and J.~A.~Wheeler, \emph{Gravitation} (W.~H.~Freeman and Company, San Francisco, 1973).

\bibitem{drell}
J.~D.~Bjorken and S.~D.~Drell, \emph{Relativistic Quantum Mechanics} (McGraw-Hill Book Company, New York, 1964).

\bibitem{chung}
W.~S.~Chung, H.~Hassanabadi and N.~Farahani,
Mod. Phys. Lett. A \textbf{34}, 1950204 (2019).


\end{thebibliography}
\end{document}